\documentclass[12pt,preprint]{aastex}

\shorttitle{Effect of Baryons on Arc/Lensed Quasar Statistics }
\shortauthors{Wambsganss, Ostriker \& Bode }

\begin{document}


\title{The effect of baryon cooling on the statistics of giant arcs and 
multiple quasars }


\author{Joachim Wambsganss }

\affil{Astronomisches Rechen-Institut, Zentrum f\"ur Astronomie
der Universit\"at Heidelberg, M\"onchhofstr. 12-14,
69120 Heidelberg, Germany}
\email{jkw@uni-hd.de} 

\and

\author{Jeremiah P. Ostriker \& Paul Bode  }

\affil{Dept. of Astrophysical Sciences, Princeton University,
Princeton, NJ 08544}
\email{jpo@astro.princeton.edu, bode@astro.princeton.edu }



\begin{abstract}
The statistics of giant arcs and large separation lensed quasars 
provide powerful constraints for the parameters of the underlying cosmological 
model. So far, most investigations have been carried out using
pure dark matter simulations.
Here we present a recipe for including the
effects of baryon cooling (i.e. large galaxy formation)
in dark matter $N$-body simulations that is
consistent with observations of massive galaxies.
Then we quantitatively compare lensing with and without applying
this baryon correction to the pure dark matter case. 
Including the baryon correction significantly
increases the frequency of giant arcs and lensed quasars,
particularly on scales of 10 arcsec and smaller:
the overall frequency of multiple images increases by about 25\%
for source redshifts between $z_s=1.5$ and $7.5$ and splittings
larger than about 3 arcsec.
The baryon rearrangement also slightly increases the fraction of
quadruple images over doubles.
\end{abstract}





\section{Introduction}   \label{sec:intro}

It has been realized for some time that 
the abundance of clusters of galaxies can 
provide very stringent constraints on cosmological models
\citep[e.g.][and references therein]{hen91,OB92,whi93,bah93,ECF96,VL96,BB03}. 
The reason for this is that--- since only 5\% to  10\% of 
the stellar mass is in rich clusters, the exact fraction 
being fixed by the definition of ``rich''--- they represent 
relatively rare fluctuations of approximately two sigma.
Consequently, as they lie on the tail of the density distribution, 
relatively small changes in the mean amplitude of the cosmic 
perturbation spectrum can produce large changes in the expected
numbers of rich clusters. 
In fact, gravitational lensing by clusters, 
which depends on the most dense fraction
of these systems, provides an especially sensitive 
measure of cosmological parameters
\citep{Turner90,1995Sci...268..274W,Bart98,LiO2002,LiO2003,LiMJLO07}.
At the present time there is significant uncertainty in the 
normalization of the fluctuation power spectrum.
For example,
Spergel et al. (2006) give an estimate based on
WMAP 3-year and SDSS data that is significantly lower
than the WMAP 1-year data, but Evrard et al. (2007) argue that the 
cluster X-ray data indicate a higher normalization cosmology. 
It is hoped that efforts such as the present paper, 
based on ray-tracing through specific cosmological models, 
can help resolve this controversy.

The frequency of giant arcs and large separation quasar lenses has
received increasing interest in the last few years
as a powerful tool for constraining cosmological parameters.  
The recent discovery of a few
widely separated quasar lenses ($\Delta \theta \ge 10$ arcsec) from the
Sloan Digital Sky Survey 
\citep{2003Natur.426..810I, 2005PASJ...57L...7I, 2006ApJ...653L..97I, 
2004ApJ...605...78O, 2005ApJ...629L..73S}
has also spurred such studies.
Almost all theoretical studies using $N$-body simulations are constrained
on dark matter only 
\citep{Bart98, 1995Sci...268..274W, 1998ApJ...494...29W,
2003MNRAS.346...67M, 2004ApJ...606L..93W, DHH, 2004ApJ...610..663O, 2005APh....24..257H,
2005ApJ...633..768H, 2005ApJ...635..795L, LiMJLO07, 
HDB07, HDBO07, 2007astro.ph..3803H}.
In such studies
it is usually assumed that the baryonic matter essentially
follows the dark matter particles. 
This may be true in general on large scales, 
but it certainly is not a very good assumption
in the central parts of galaxy clusters or for very massive
isolated galaxies, which are precisely the
essential matter concentrations for the production of giant arcs and
wide quasar lenses.
The baryonic component is also known to dominate in the central regions
of elliptical galaxies.

Recent studies have begun investigating the influence of baryons 
on DM clustering.
Puchwein et al. (2005) looked into the effect of gas physics on
strong lensing by individual galaxy clusters. They found that cooling and 
star formation can increase the strong-lensing efficiency 
considerably.
Jing et al. (2006) studied the influence of baryons on the clustering
of matter and weak-lensing surveys, and found that the clustering of
total matter is suppressed  by about 1\% on large
scales of $ 1 h {\rm Mpc}^{-1} \la k \la 10 h {\rm Mpc}^{-1}$, 
while it is
boosted between 2\% and 10\% on small scales of 
$ k \approx 20 h \rm {Mpc}^{-1}$; they conclude that this
should be measurable with future weak lensing surveys
(Jing et al. 2006).
Lin et al. (2006) looked into the influence of baryons on the 
mass distribution of dark matter halos and found an increase of
the concentration parameters by 3\% to 10\% as compared to pure
$N$-body simulations. 
Rozo et al. (2006) studied the effect 
of baryonic cooling on giant arc abundances for individual clusters;
they found that the arc abundances can 
be increased by factors of a few.

In this study we quantitatively 
investigate the effect of baryon cooling on the 
statistics of arcs and widely separated multiple quasars.
We first motivate and describe a simple recipe for redistributing part
of the matter in the centers of halos to approximate the
effects of cooling and star formation,
and present some tests of this method. Subsequently we
summarize our use of the ray shooting method. 
Then we describe the quantitative
results  of our baryon redistribution recipe by directly
comparing pseudo-three-dimensional
ray shooting simulations
with and without this baryon cooling, and finally
discuss this effect with respect to the observational situation.

\section{Methods}   \label{sec:methods}

\subsection{Baryon rearrangement}   \label{sec:br}

\begin{figure}
\plotone{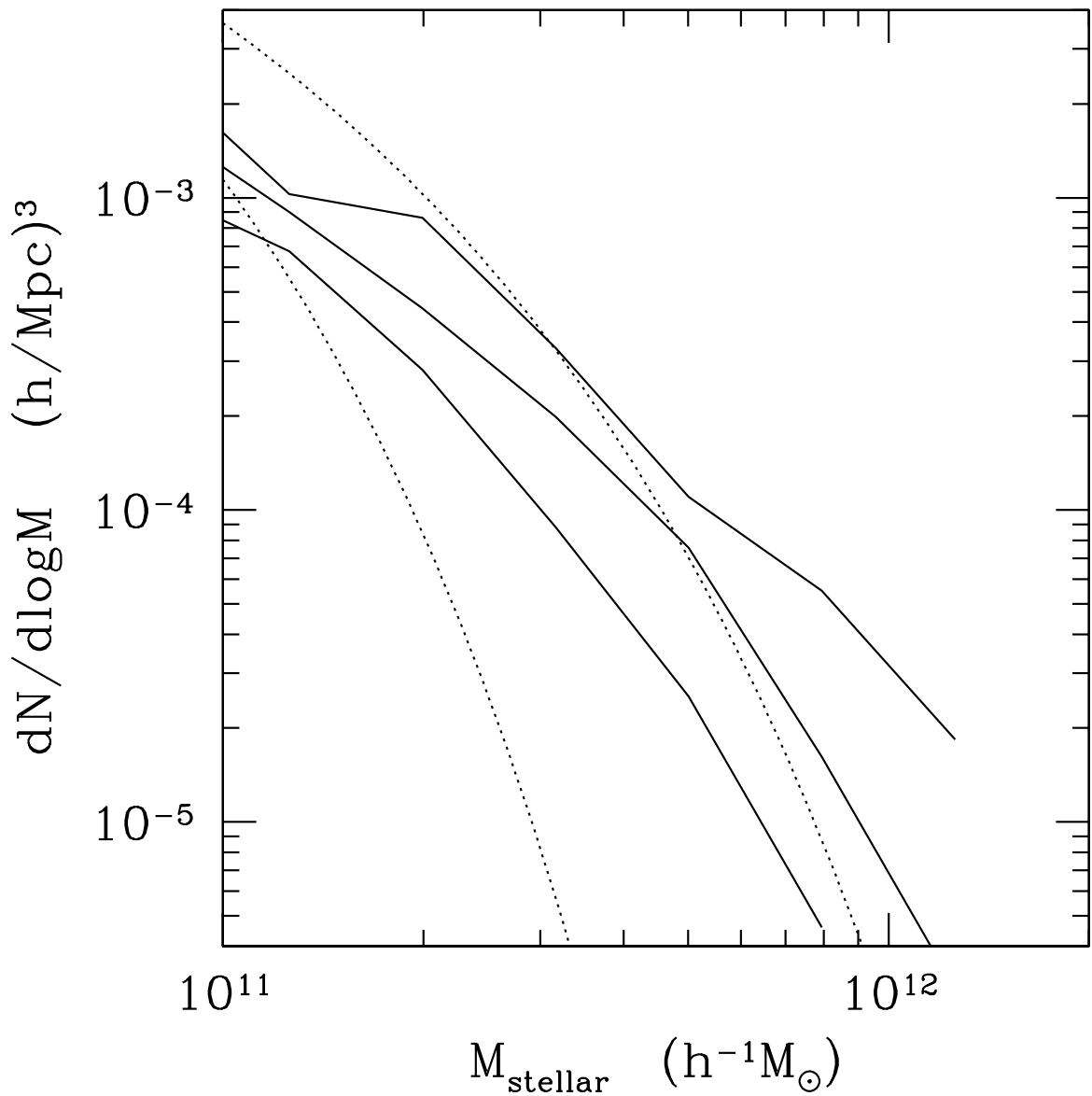}
\caption{Solid lines: stellar mass function as determined with 
the recipe described in the text, at  redshifts 
$z$=0.05, 0.55, and 1.05 (top to bottom).
The lower dotted line is taken from the SDSS luminosity function
(Blanton et al. 2003, Gavazzi et al. 2007), assuming
$M_{star}/L=4.4$; the upper dotted line instead assumes 
$M_{star}/L\propto L^{0.25}$.
\label{fig:allms} }
\end{figure}

\begin{figure}
\plotone{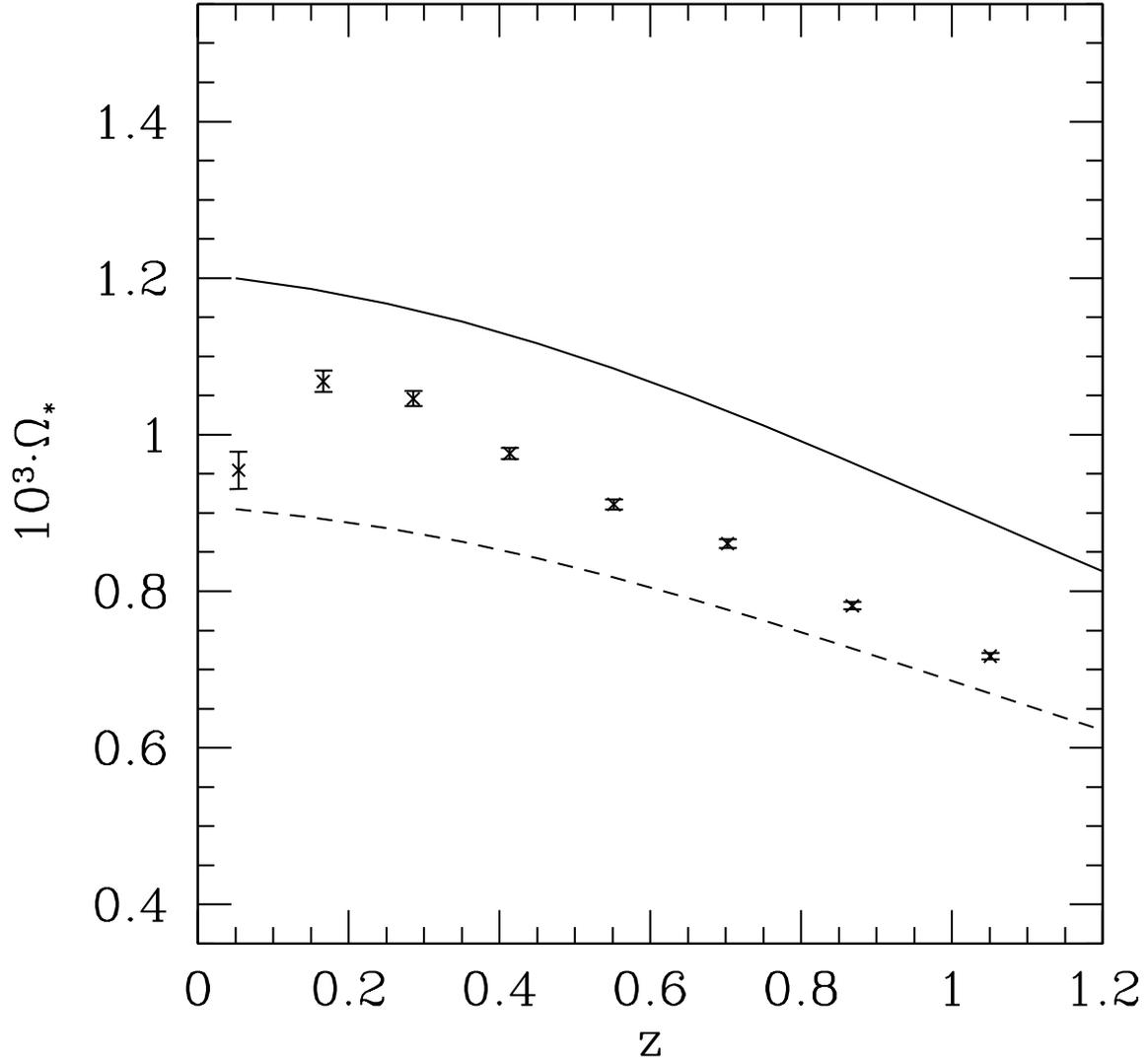}
\caption{The total rearranged ``stellar''
mass as a function of redshift (determined
by applying the baryon redistribution recipe described in
Section 2.1).  The solid line
is an  empirically based estimate of
the total mass in elliptical galaxies from 
\citet{Nagamine+2006};  the dashed line
is limited to systems with
$V_c>215$km/s, which would produce a splitting angle
of $>$3''.  
\label{fig:stellarmass} }
\end{figure}

\begin{figure}
\plotone{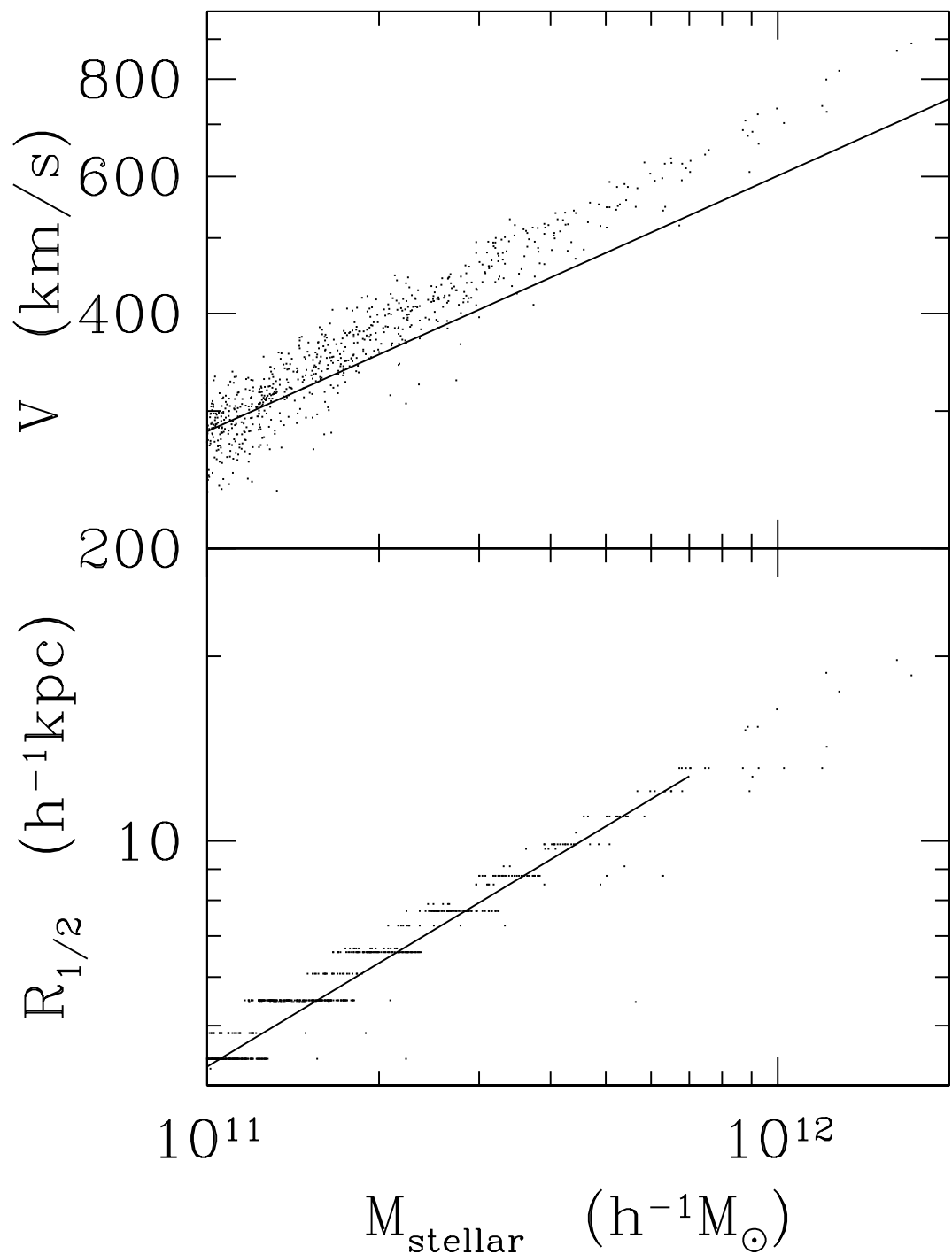}
\caption{Top panel: Circular velocity $V$ at $R_A$ 
as a function of stellar mass $M_{\mathrm stellar}$, 
for halos identified in the $z$=0.05 and 0.17 lens planes.  
The line is the circular velocity measured at 2.2 disk lengths by 
\citet{Pizagno05}.
Bottom panel: Half-mass radius $R_{1/2}$
of the redistributed mass 
as a function of stellar mass $M_{\mathrm stellar}$.
The line
is the Sersic half-light radius of early type galaxies
from SDSS, see \citet{Shenea2003}.
\label{fig:mvr} }
\end{figure}

In order to approximate the effects of galaxy formation in
our ray shooting simulations, 
we locate halos in all lens planes and identify the amount of mass
likely to have cooled into stars.  
This mass will be
rearranged such that the inner part of the total profile is isothermal.
Such an approach is supported by observations, e.g.
\citet{Peng+04} or 
\citet{Gavazzi+07}, who have found that early-type galaxies
are consistent with an isothermal profile out to $\ge 300$ $h^{-1}$kpc
from both kinematical and lensing data.
Here we describe in detail the ways this rearrangement is performed.
The approach that we adopt  is admittedly very rough, and so, 
as in all semi-analytical modeling, we constrain the free
parameters to respect observational data. 
The requirements the model must fulfill are:
a) The fraction of the baryonic component which is rearranged
into stellar systems is consistent with the stellar mass fraction of
the universe 
(cf. Figures \ref{fig:allms} and \ref{fig:stellarmass}).
b) The cosmic buildup of mass in stellar systems parallels what is 
known from observations 
(Figure \ref{fig:stellarmass}).
c) Individual mass profiles within the baryonically dominated systems 
are consistent with kinematic and gravitational lensing data
(Figure \ref{fig:mvr}).
d) The distribution of systems as a function of stellar mass and
circular velocity approximates observational data
(Figures \ref{fig:allms} and \ref{fig:mvr}).

Suppose we have projected the mass in some volume
onto a plane for the purpose of ray tracing (as
described in the next section).
Let us denote the surface density of a given pixel
as $\sigma$ and the pixel size as $l$, so that the
mass in each pixel is $\sigma l^2$.
For a given 2-D plane, each pixel is examined in turn to see
if it has a higher surface density than any other
pixel within a radius roughly the size of the smallest
objects resolved in the simulation, 
$R_b = 30 h^{-1}$kpc physical.
If so, this pixel is taken to be the center of a halo.
For a given cooling radius $R_c$ (to be 
determined below), the background surface
density $\sigma_{bg}$ is set to the mean surface density
between $R_c$ and $R_c+(R_b/2)$.  
The mass of the halo is defined as
\begin{equation} \label{eqn:masshalo}
M_h = \int W(R/R_c)\left[\sigma({\bf R}) - \sigma_{bg} \right] dA
\hspace{1cm} ,
\end{equation}
integrating from the halo center out to $R_c$;
here we will use the window
function $W(x)=[1-x^2]^2$.
With the mass weighted radius of the halo given by
\begin{equation} \label{eqn:radhalo}
R_h = M_h^{-1} \int R W(R/R_c)
       \left[\sigma({\bf R}) - \sigma_{bg} \right] dA
\hspace{1cm} ,
\end{equation}
the halo temperature is defined as
\begin{equation} \label{eqn:thalo}
T_h = \left( \frac{Gm_p}{k} \right) \frac{M_h}{R_h}  
\hspace{1cm} ,
\end{equation}
with $G$ and $k$ being the Newton and Boltzmann constants,
respectively, and $m_p$ the proton mass.

We wish to find a cooling radius such that half the baryons
contained within this radius will have formed stars.  For ionized
gas in a galaxy cluster the cooling time from
Bremsstrahlung radiation is given by 
$t_{br} = 9\times 10^7 T_8^{1/2} n_e^{-1}$yr 
\citep[cf.][]{AllensAQ};
here $n_e$ is the electron density per cm$^3$,
which we will take to be $n_e = 3M_h/(4\pi R_h^3 m_p)$.
Let $T_0$ be the temperature at which the cooling time
equals the age of the universe at the redshift under
consideration, $t_{hub}$:
\begin{equation} \label{eqn:t0}
T_0 = n_e^2 \left( \frac{t_{hub}}{9\times 10^7 \rm{yr}} \right)^2 10^8 \rm{K}
\hspace{1cm} .
\end{equation}
To set $R_c$,
we start with a small estimate for the cooling radius, and increase
it until
\begin{equation} \label{eqn:fcirt}
e^{-T_h/\eta T_0} = f_*
\hspace{1cm} ,
\end{equation}
with the cooled fraction $f_*=\onehalf$,
is satisfied.  
A number of oversimplifications have gone into the calculation
of $T_h$ and $T_0$, so the variable $\eta$ is introduced to 
account for these approximations.  This parameter can be adjusted
to give more or less stellar mass; we find $\eta=\onethird$ gives
a reasonable amount of stellar mass, consistent with observational
constraints.
Problems will arise with this algorithm if the aperture $R_c$
increases to the point where it begins to include neighboring
structures, as can happen in the case of a galaxy-sized halo
inside a group or cluster.  Thus, if during this process it 
happens that $R_h/R_c>\onethird$, we stop increasing $R_c$
and leave $f_*$ larger than a half.

\begin{figure}
\plotone{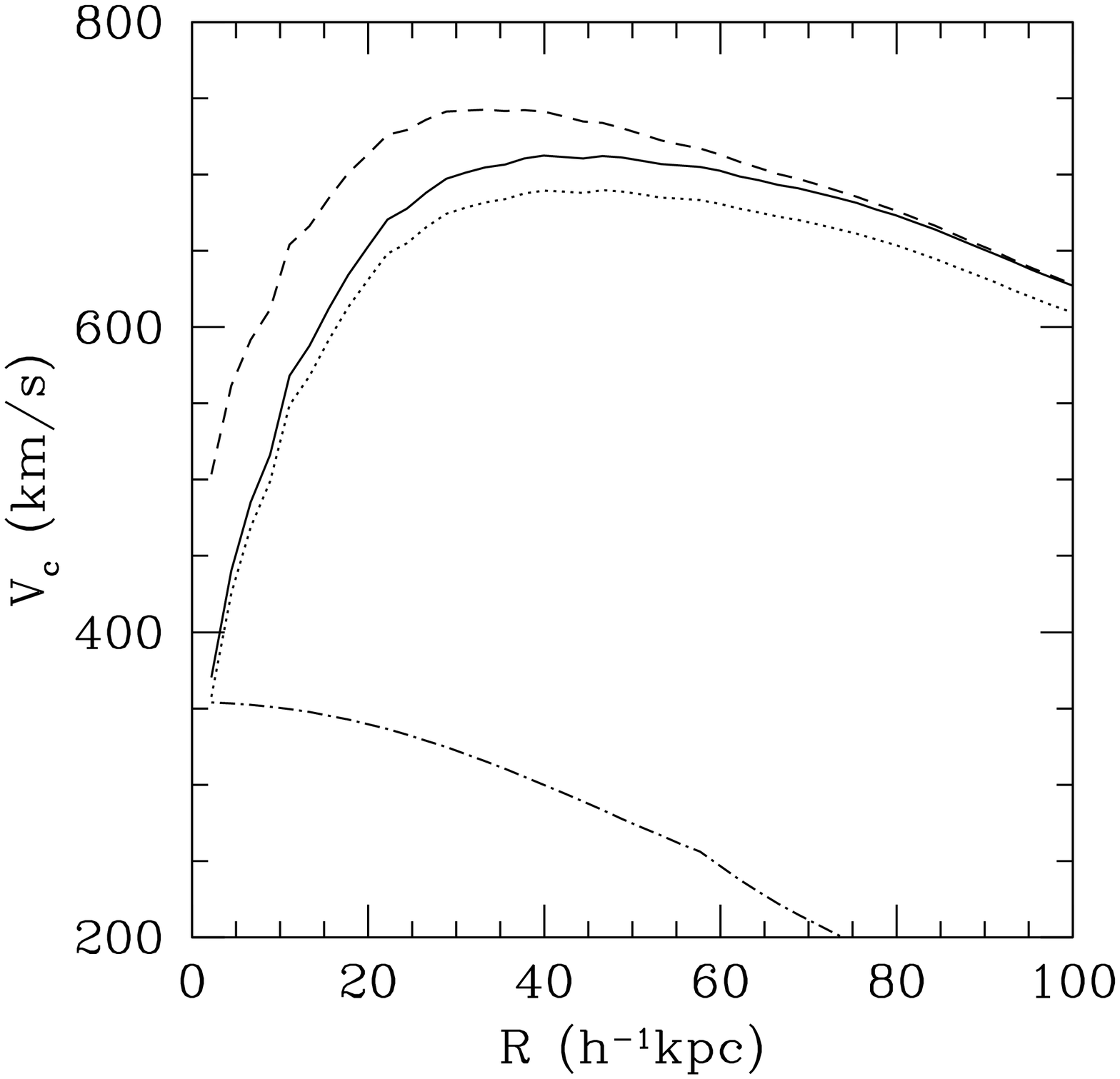}
\caption{The circular velocity profile of an example halo, demonstrating
the effect of redistribution.  
Solid line: initial circular velocity.  Dotted line: after mass removal
(cf. Eqn. \ref{eqn:sint}).  Dot-dashed line: velocity of
the isothermal component (cf. Eqn. \ref{eqn:sfin}).
Dashed line: the final circular velocity profile.
\label{fig:vexample} }
\end{figure}

Mass representing the cooled baryonic component
is then removed from inside $R_c$, reducing the surface density
to a temporary value 
\begin{equation} \label{eqn:sint}
\sigma_i({\bf R}) = \sigma({\bf R}) - \frac{\Omega_b}{\Omega_m} f_* W(R/R_c) 
  \left[ \sigma({\bf R}) - \sigma_{bg} \right]
\hspace{1cm} ,
\end{equation}
with the caveat that any pixels denser than the central pixel 
are left unchanged.
This removed mass is then added back in, using an isothermal 
profile $A/R$ within a smaller radius $R_A$.
With the isothermal component added back in,
the final surface density is
\begin{equation} \label{eqn:sfin}
\sigma_f({\bf R}) = \sigma_{bg}+ A W(R/R_A)/R 
\hspace{1cm} .
\end{equation}
For a given $R_A$ (set below)
the value of the constant $A$ is set by conservation of mass,
i.e. the mass put back in, $M_s$, must of course
equal that taken out originally.
Figure \ref{fig:vexample} illustrates how our method of baryon 
redistribution affects the velocity profile of a halo. 
The four lines in this example halo show the initial circular 
velocity from the pure N-body simulation (the solid line), 
the circular velocity after mass removal as given by 
Eq. \ref{eqn:sint} (dotted), the velocity curve of the 
isothermal component to be added back in according 
to Eq. \ref{eqn:sfin} (dot-dashed), 
and the final circular velocity profile after the redistribution 
is complete (dashed line).  This example 
corresponds to a brightest cluster galaxy residing at 
the center of the cluster dark matter halo.

Figure\,\ref{fig:allms} shows the ``stellar mass function''
(that is, the distribution of $M_s$)
in halos identified by our algorithm at the 
redshifts $z = 0.05, 0,55, 1.05$. 
The distribution of stellar masses is quite similar at
all three redshifts.  The lower dotted line results from taking
the luminosity function of SDSS galaxies \citep{Blantonea2003}
and assuming a constant $M_{star}/L=4.4$ 
\citep[see][]{Gavazzi+07};
the upper dotted line instead assumes $M_{star}/L\propto L^{0.25}$
\citep{VO2006}.
As a further test,
summing the total mass rearranged at a given redshift
yields an estimate of the stellar mass in large spheroidal galaxies.
The points in Figure \ref{fig:stellarmass} show the density
of this component (as a fraction of the critical density)
versus redshift.  The error bars give the error on the mean
value of the 486 planes used; this error becomes
larger at low redshift
primarily because the size the planes is smaller.
The total fraction of mass that is rearranged rises 
slowly from $\Omega_{star}=0.7\times10^{-3}$ at $z=1$ to
$\Omega_{star}=1.0\times10^{-3}$ at $z=0.29$, remaining
at this value for lower redshifts.
Also shown are  empirically
based estimates \citep[compiled in][]{Nagamine+2006} for
the total mass in elliptical galaxies.  The solid line shows
the total mass in elliptical 
(``spheroidal'') components, and the dotted line indicates
an estimate of the mass limited to systems with
$V_c>215$km/s which would produce a splitting angle
of $>$3 arcseconds.   These lines bracket our points, indicating
that we are not overproducing galaxies, but are including
those likely to produce significant lensing.
From these tests we conclude that our choice of $\eta$,
which determines the mass rearranged in a given halo, is reasonable. 
The stellar mass function is insensitive to this
parameter--- doubling $\eta$ causes very little change
to the mass function above $2\times 10^{10}h^{-1}M_\odot$.

The value of $R_A$ is set to match the observed
size distribution of SDSS galaxies.  \citet{Shenea2003},
using the stellar masses of \citet{Kauffmannea2003}, find
that the Sersic half-light radius of early type galaxies
varies with stellar mass as $M^{0.56}$.   
When using the above procedure on the lowest redshift planes,
we find that the cooling radius $R_c$ varies with the mass
removed as $R_c \propto M_s^{0.20}$.  Thus by setting
\begin{equation} \label{eqn:rafin}
R_A = R_c \left( \frac{M_s}{2.4\times 10^{13} h^{-1}M_\odot} \right)^{0.36} \ \ \ \ ,
\end{equation}
the half-mass radius of the added mass matches very well the 
half-light versus radius relation for early-type
galaxies of \citet{Shenea2003};  this is shown at low redshifts in the
bottom panel of Figure \ref{fig:mvr}.
If this choice of $R_A$ would result in mass being added 
to a pixel which is below the background density, 
i.e. $\sigma_i  <  \sigma_{bg}$, we instead reduce $R_A$ to
the level where this will not occur; it is quite rare that
this adjustment is invoked.

The value of $R_A$ also sets the isothermal circular 
velocity $V_c=2\sqrt{GA}$.   The upper panel of Figure \ref{fig:mvr} 
displays the circular velocity measured at $R_A$ as a function
of stellar mass.
These velocities  agree  reasonably well with the relation between 
stellar mass and rotation 
velocity at 2.2 disk lengths measured by \citet{Pizagno05},
who find that $V_{2.2}$ increases
with stellar mass roughly as $M_s^{1/3}$.
At higher stellar masses our circular velocity is about
10\% higher than the measured $V_{2.2}$.
A lower limit on the isothermal circular velocity
is imposed: for $V_c<125$km/s, the halo is simply returned to
its original density profile.
This velocity cut also puts a lower limit on the splitting
angles affected by this procedure, which is of order one arcsecond.

To summarize this section, the simple procedure described here
reproduces the stellar mass function of large galaxies as
a function of redshift, with
the appropriate sizes and circular velocities.   In Section 3
we investigate what effects these galaxies with rearranged 
matter distribution have quantitatively on the lensing frequency.

\subsection{Mass planes, ray shooting and identification of arcs}   \label{sec:planes}

The methods used to create the mass distribution
in the lensing planes and to carry out the ray tracing
are described in detail in 
\citet{2004ApJ...606L..93W}. 
Here we briefly summarize the procedure and give the numerical
values of our parameters. Essentially, light rays are followed
backward from the observer through a series of lens planes, 
approximating a  three-dimensional matter distribution;  the
lens planes are obtained from $N$-body simulations. 
The parameters of the underlying cosmological model are: 
$\Omega_{\mathrm M}=0.3$, 
$\Omega_\Lambda=0.7$, 
$H_0=70$ km/sec/Mpc, 
$\sigma_8$=0.95,
and $n_s$=1.
The $N$-body simulation was performed in 
a box with a comoving side length of  
$L=320 h^{-1}$Mpc.
We used $N=1024^3$ particles,
so the individual particle mass is
$m_{\rm p}=2.54\times 10^9 h^{-1}$ M$_\odot$.
The cubic spline softening length for all
particles was set to $\epsilon=3.2 h^{-1}$ kpc.
The output was stored at 19 redshift values out to 
$z \approx 6.4$, such that the centers of the saved
boxes matched comoving distances of 
$(160 + n \times 320) h^{-1}$Mpc, where $n=0,...,18$. 

For 243 lines-of-sight traversing the box,
two lens planes were produced by
bisecting along the line-of-sight and projecting the mass
in each $160 h^{-1}$Mpc--long volume onto a plane.
In the lowest redshift box the lensing
planes are $1.9 h^{-1}$Mpc on a side, and the pixel size 
is $l=2.3 h^{-1}$kpc.  
With increasing lens
redshift,
we keep the number of pixels constant but increase the
physical size of the planes, such that the pixel opening angle 
remains constant.  

Light rays are then propagated backwards through these lens planes,
beginning with a regular grid at the lowest redshift lens plane
(i.e. the image plane) and working to higher redshift by
considering the proper deflection in each lens plane.
For each source redshift, the coordinates of the rays in the
image plane and in the source plane are stored. For analyzing the
imaging properties, we impose
a regular grid of sources in the source plane and identify
for each source position the image multiplicity 
(by far most of them are single images, some are
triples, we also identify quintuples, septuples etc.), and
then for each image the corresponding image position
and image magnification. 

In the literature, different authors use different definitions 
for what an ``arc'' is.
As stated in Wambsganss et al. (2004), we demand as a miminum 
requirement that an arc be part of a multiple image system 
{\em and} have a certain minimum magnification (we chose different
values between 5 and 25 as threshholds). Since 
strongly lensed sources are almost always highly distorted in the 
tangential direction, in particular the strongly magnified
multiply imaged ones, 
the magnification is in general
close to the length-to-width ratio of such an image.  
We double-checked this assumption in a number of individual
cases and found good agreement above the 90\% level.

For the multiply imaged sources
we order the images by magnification and determine the distances
between the images, whereby the ``image separation'' is defined
as the angular distance between the brightest and the third
brightest image. 
More details are given in 
Wambsganss et al. (2004, 2005).

\section{Ray shooting Results and comparison with observations}   
	\label{sec:rayshoot}

\subsection{Ray shooting }   \label{subsec:rayshoot}
We performed two complete sets of rayshooting simulations 
as detailed in Subsection  2.2,
once with the original pure $N$-body matter screens
and once with the new recipe of rearranged matter, 
creating 100 independent realizations for each of the two.
This procedure allowed us to study the effect 
both statistically and on individual lines-of-sight 
(the latter turned out to be essential for a complete 
understanding of the results, as described below).
Three source redshifts are
considered, $z_s=$ 1.5, 3.7, and 7.5.

\begin{figure}
\plotone{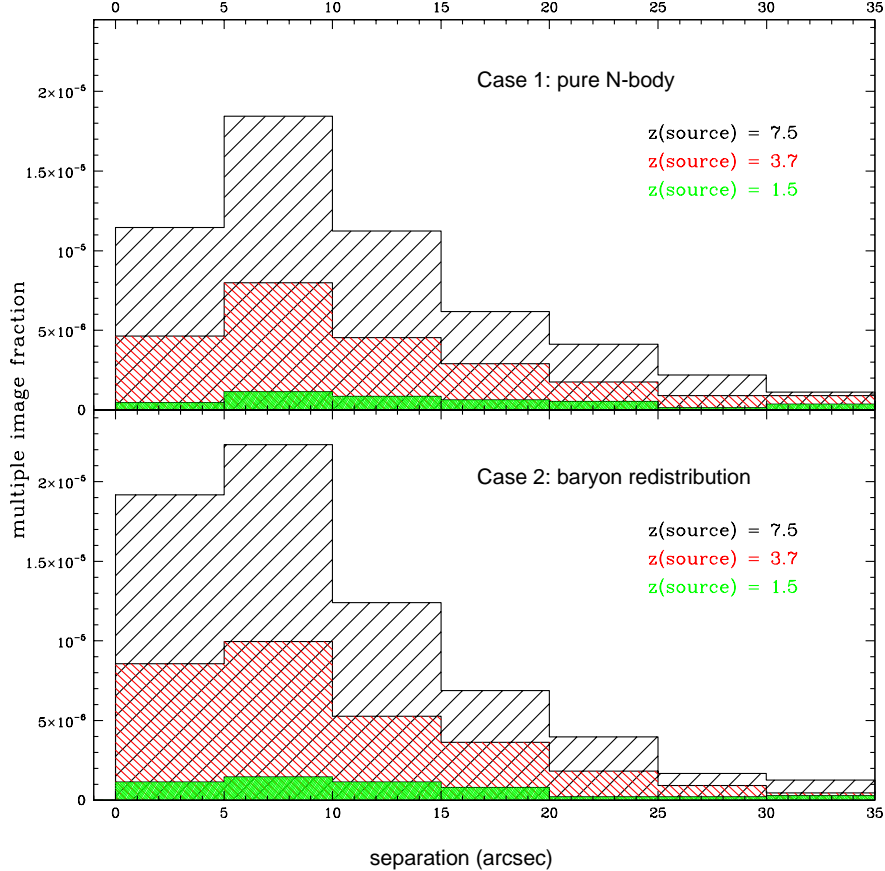}
\caption{Frequency of multiple images as a function
of separation (binned with 5 arcsec width). 
The number of multiple images clearly
increases with decreasing splitting angle; the drop in
the smallest separation bin is due to the fact that we are
highly incomplete for splitting angles below 3 arcseconds.
The N-body case is shown in the top panel, the scenario with
baryon redistribution in the bottom panel. The histograms reflect
three different source redshifts: 
z$_s$ = 7.5 (highest, black, loose shading), 
z$_s$ = 3.7 (middle, red, intermediate  shading), 
z$_s$ = 1.5 (lowest, green, densest shading).
In the baryon redistribution scenario we find 
more multiple images (on average about 25\%); 
this is most pronounced in the small separation regime. 
\label{fig:splitting} }
\end{figure}

\begin{figure}
\epsscale{.49}
\plotone {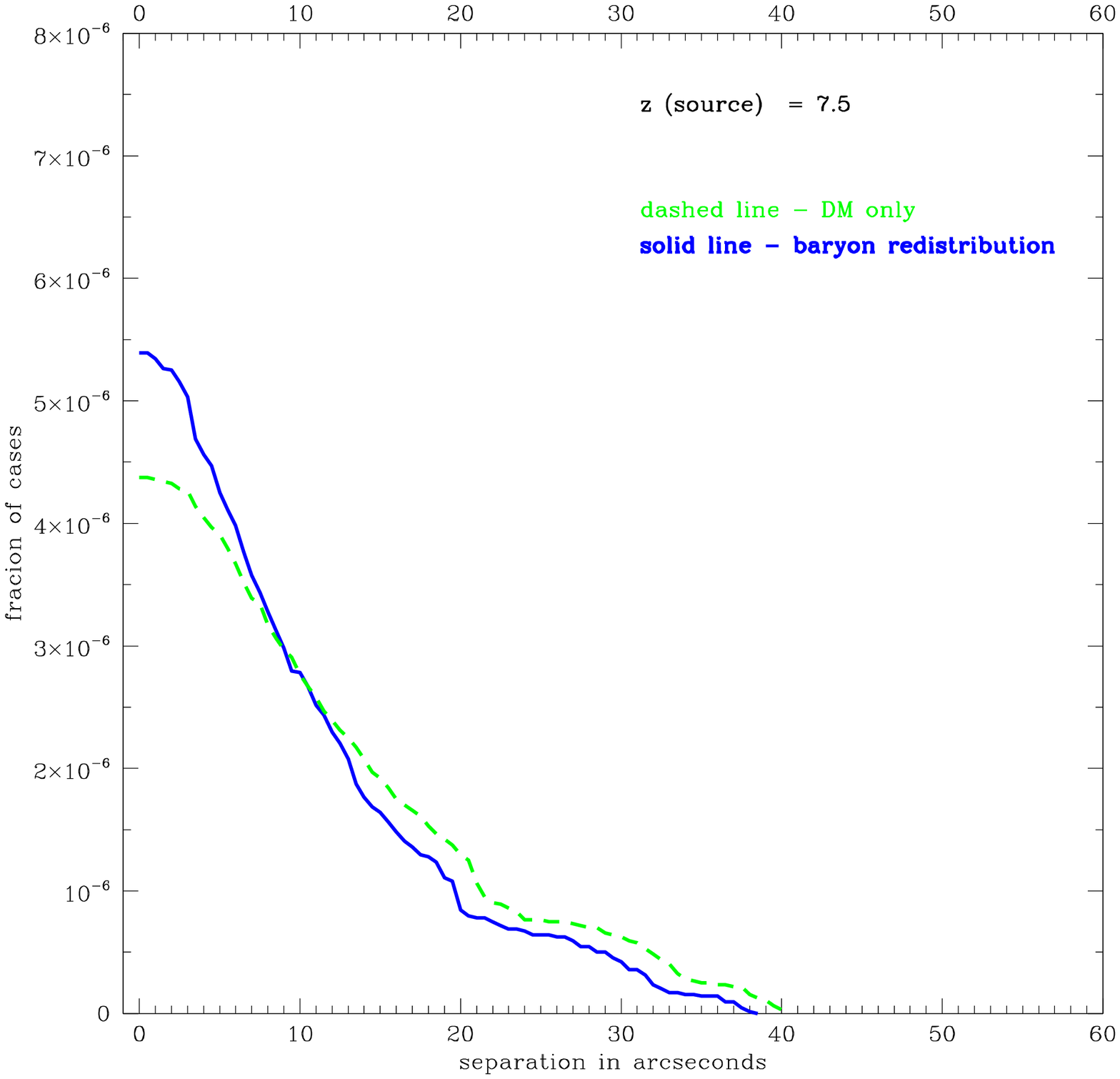} 
\plotone {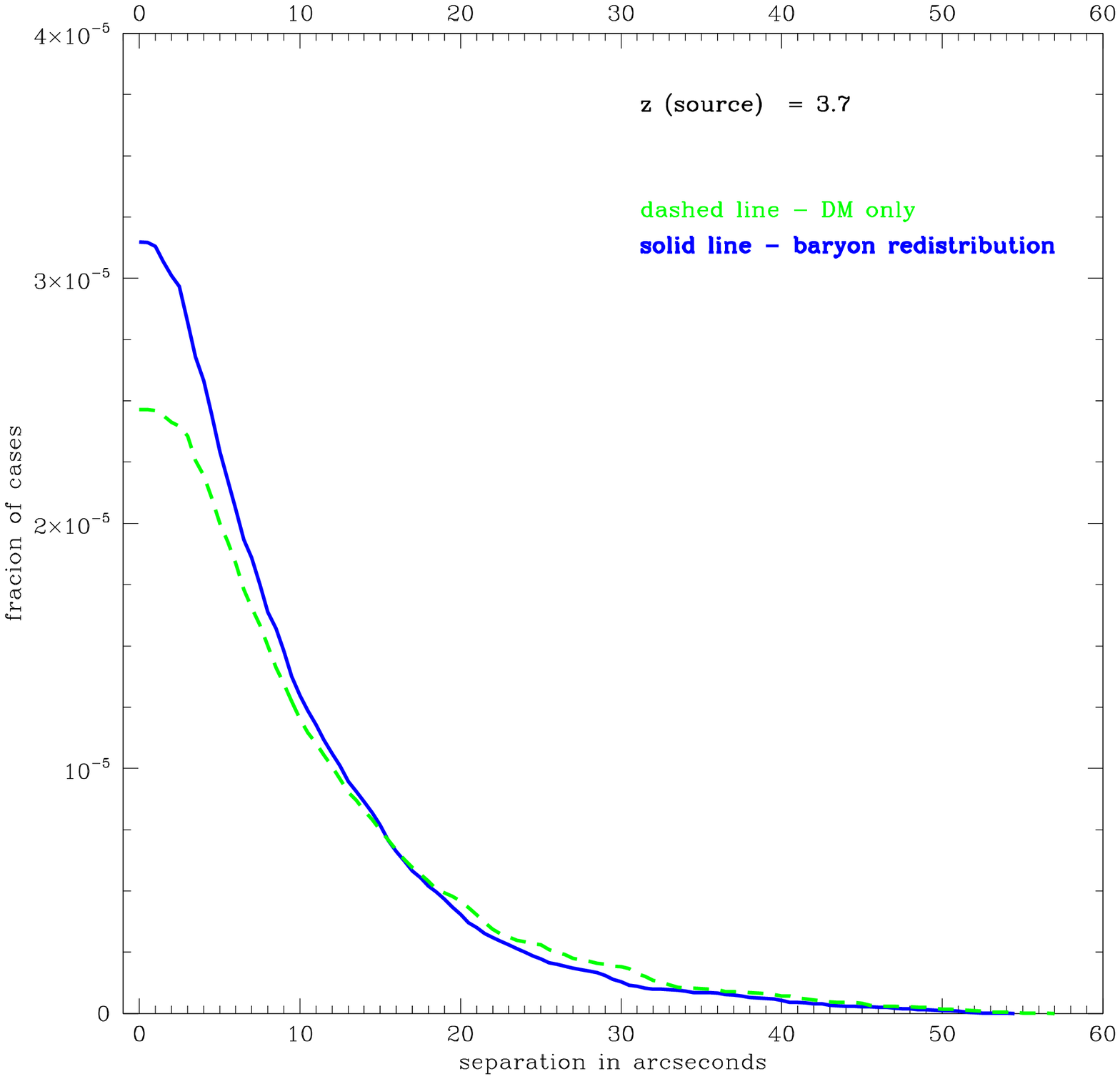} 
\plotone {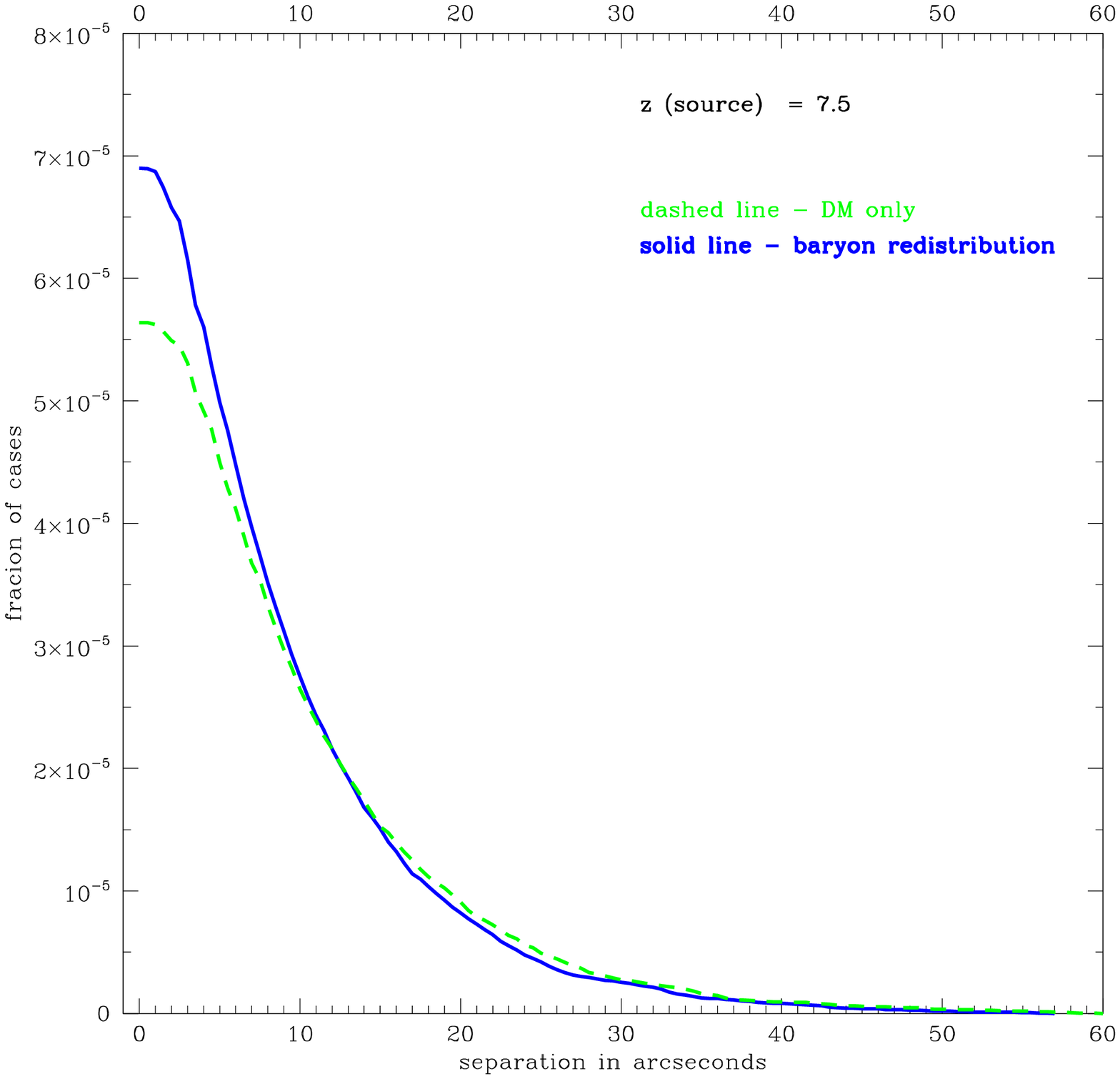} 
\caption{Integrated probability distribution of multiple images as a 
function of separation for
a dark matter only simulation (green dashed line) and 
with baryon redistribution as described in the text (blue  solid line).
Top panel: For a source redshift of $z_s = 1.5$;
middle panel: $z_s = 3.7$; 
bottom panel: $z_s = 7.5$.
\label{fig:lenspdf}}
\end{figure}

With the rearrangement,
the number of multiple images increased for all source redshifts,
particularly for angular
separations up to 15 arcsec. 
This is shown in 
Figure \ref{fig:splitting}; where
the multiple image cases are binned according to
the image splitting using bin widths of 5 arcsec 
(note
we are highly incomplete in the first bin, because we do not properly
resolve image splittings smaller than about 3 arcseconds).
The level of increase can be read off of the integrated
distributions in 
Figure \ref{fig:lenspdf}, where the 
dashed lines show the integrated frequency of multiple images 
as a function of separation for the pure N-body case, and the solid
lines show the scenario with baryon rearrangement.
Qualitatively, it is obvious
that including the
baryon redistribution produces more multiple images.
The amount by which the rearranged case is higher than the pure $N$-body
case is almost independent of source redshift: 
the fraction is about 25\%
for the redshift range $z_s = 1.5$ to $7.5$
(it varies between 22\% and 28\%).
Beyond the total lensing frequencies,
it is interesting to study the 
{\em differential} effect: it is
for relatively small separations ($\le 10$ arcmin) in particular 
that the baryon-redistribution case
leads to more multiple images:
about a 70\% increase in the lowest bin, and 30\%
increase in the 5--10 arcsecond bin.
This excess decreases with
increasing separation, such that
at a splitting of $\sim 12$ to $15$ arcsec the two cases result
in the same frequencies for multiple images. 
This effect is easy to understand: 
the baryon redistribution steepens the innermost parts of the 
density profiles of halos, and hence can bring DM
halos which are originally just below the
critical density for lensing above this value, 
thus producing multiple images. 

\begin{figure}
\plotone{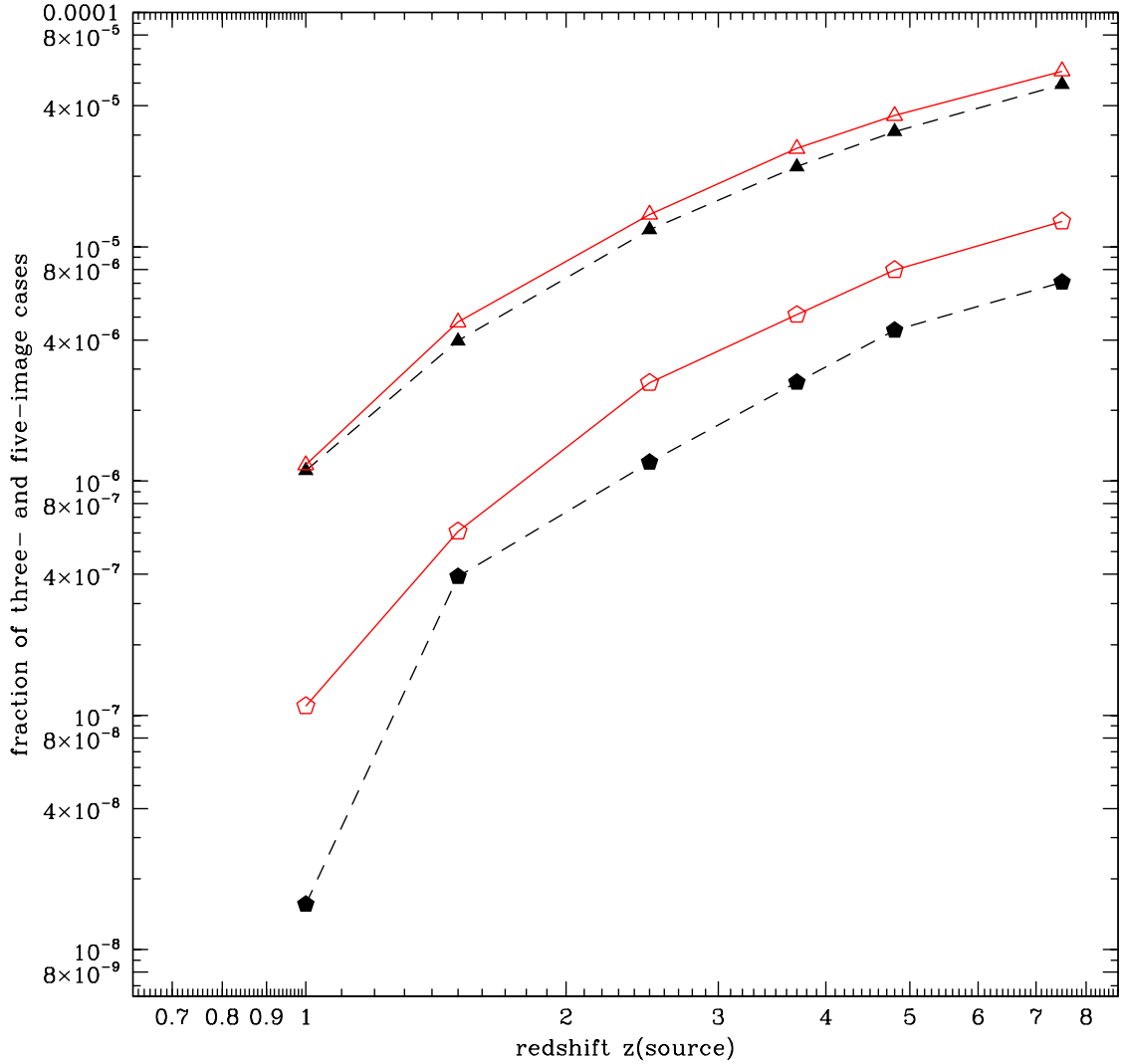}
\caption{Image multiplicity as a function of source redshift for
"triple" images (triangles) and "quintuple" images (pentagons). 
The pure N-body case is shown with solid symbols and dashed lines, 
for the baryon-rearrangement  case we used open symbols and solid lines.
\label{fig:multi} }
\end{figure}

At still larger separations, the
baryon redistribution case apparently produces slightly fewer cases
(at the few percent level).  
The reason for this
is not quite so obvious; in fact, it is rather counter-intuitive:
if a fraction of the total mass becomes more concentrated,
the total mass inside an image pair should be the same or
larger.  The explanation is that this effect
is just an artifact of our definition of image separation. 
As described above, 
we define the separation of a multiple image system 
as the angular distance between
the brightest and the third-brightest image. 
Our method always identifies
the ``odd'' image in triples and quintuples,
although in practice most would
be identified as double and quadruple images, because the
odd images in most observational situations are demagnified and not
detectable (hence we refer subsequently to those cases as 
"doubles" and "quadruples").
A detailed look at the distribution of multiple images shows that 
a side effect of the redistribution of
baryons is that the number of quadruple images increases more strongly
than the number of doubles 
(see Figure \ref{fig:multi}). 
We are able to verify this by comparing individual lines of sight, 
with and without the matter rearrangement.
When we compare one-to-one a situation
that produces a double image in the DM-only scenario 
and a quadruple image in the
baryon-redistribution case, the additional image pair occasionally is
brighter than the faint ``odd'' image, and it happens that
this pair is close to the
brightest image of the multiplet, resulting in a smaller 
measured separation than in the pure N-body case.
This leads to the apparent
small deficit of large separation cases in the redistributed-baryon case.

\subsection{Comparison with observations}   \label{subsec:obs}
Over the last decade, various papers have investigated the
observational occurrence of giant arcs statistically, e.g.
\citet{L99}, 
\citet{Gladders+03},
and \citet{Sand+05}. 
The frequency of arcs reported in these studies
varies from one giant arc
per 45 deg$^2$ to about one arc per 10 deg$^2$; the large spread 
may be explained partly by the slightly different definition 
and limiting magnitude, and may 
be related to the relatively small number
of cases per study.
Overall, just about 50 arc systems had been used for these 
statistical analyses, found in various surveys
with diverse selection criteria.

Two recent studies based on SDSS data present quite different
results as well.
On one hand,  \cite{Estrada+07} report no arcs found in a
systematic investigation of 825 SDSS clusters (other than
one serendipitous discovery). 
On the other hand, \cite{Hennawi+06} presented
first results from a new survey for giant arcs, which 
may as much as double the  number of known arcs 
(about 30 new systems already reported). This study
is very promising not just due to the large number of new
arc systems to be expected, 
but -- at least as importantly -- also by the fact
that it will find them by well defined selection
criteria. 

Until these future results will be published,
we can compare our results here 
(although only ``differentially'')
to those in \citet{2004ApJ...606L..93W}. 
There it was found that the simulated
pure N-body based arc frequencies are consistent with the
observations, if we allow the source redshifts 
to extend well beyond $z \ge 1$ (as the distribution of
observed arcs is as well).  The result presented here---
that by including the baryon rearrangement the 
predicted number of arcs increases by about 25\%---
puts the predicted arc frequency a bit on the high side, 
though still with considerable uncertainty
both on the observational (e.g., absolute numbers, selection criteria) 
and the theoretical side (e.g., 
parameters of underlying cosmological model, exact definition
of ``arc'').  
Furthermore, the model used here has a high normalization
for the amplitude of fluctuations as compared to the recent
results from WMAP; a lower normalization  would significantly
reduce the lensing frequency \citep{LiMJLO07}.

\section{Summary and Conclusion}   \label{sec:summary}

This paper quantitatively investigates the effects on lensing of 
baryon redistribution in originally pure dark matter $N$-body simulations. 
We describe a heuristic recipe for rearranging the baryons in
dense dark matter halos that is consistent with
observational data for massive galaxies, 
present a number of tests of this prescription,
and then apply it
to a cosmological simulation used for 
multiple lens plane ray shooting.
We compare the frequency of multiple images, the image separation
and the image multiplicity between the 
original $N$-body matter distribution and the redistributed version.
We find that on average the case 
taking into account the redistribution of baryons
produces 25\% more multiple images;
this is almost independent of the source redshift in 
the range $z_s =$ 1.5 to 7.5.  
For splittings between 5--10 arcseconds, the number of multiple images
increases by roughly 30\%, and by 70\% for smaller separations $<5$
arcsecond.
But this last result must be treated with caution since we
are resolution limited for splittings  $\lesssim 3$ arcsecond.
As noted, most of the new multiple images
systems occur for angular scales $\le 10$ arcseconds.
We also find that the number of quadruple images increases more than the
number of double images. This produces an apparent slight reduction of 
larger separation cases, which is in fact only an artifact of the
way we define image separation, namely as  the distance between 
the brightest and the third-brightest image.
Since arc statistics are such a good tool for
distinguishing between models with different cosmological
parameters, we would like to  emphasize once again
the need for both very good observational studies and 
for more explorations of the important parameters in the
simulations.




\acknowledgments

This work was supported by the European Community's Sixth Framework
Marie Curie Research Training Network Programme, Contract No.
MRTN-CT-2004-505183 ``ANGLES''.
This research was also facilitated through an allocation of advanced 
computing resources from the National Center for Supercomputing 
Applications (under grant MCA04N002) and the Pittsburgh 
Supercomputing Center.  In addition, computational facilities at 
Princeton supported by NSF grant AST-0216105 were used, as well as
high performance computational facilities supported by 
Princeton University under the auspices of the
Princeton Institute for Computational Science and Engineering 
(PICSciE) and the Office of Information Technology (OIT).

\clearpage



\epsscale{.95}

\epsscale{.90}

\epsscale{.75}

\epsscale{.95}

\end{document}